\documentclass[10pt, aps, prl, twocolumn, showpacs, floatfix]{revtex4}
\usepackage{amsmath, amssymb, mathtools, array, color, bbm}
\usepackage{hyperref}
\hypersetup{colorlinks=true, linkcolor=blue, citecolor=blue}
\usepackage{graphicx}
\usepackage{times}

\begin{document}


\title{Quantum synchronization of a driven self-sustained oscillator}

\author{Stefan~Walter}
\author{Andreas~Nunnenkamp}
\author{Christoph~Bruder}
\affiliation{Department of Physics, University of Basel,
Klingelbergstrasse 82, CH-4056 Basel, Switzerland}

\date{\today}


\begin{abstract}
  Synchronization is a universal phenomenon that is important both in
  fundamental studies and in technical applications. Here we
  investigate synchronization in the simplest quantum-mechanical
  scenario possible, i.e., a quantum-mechanical self-sustained
  oscillator coupled to an external harmonic drive.  Using the power
  spectrum we analyze synchronization in terms of frequency entrainment
  and frequency locking in close analogy to the classical case.
  We show that there is a step-like crossover to a synchronized state
  as a function of the driving strength. In contrast to the classical
  case, there is a finite threshold value in driving. Quantum noise reduces the
  synchronized region and leads to a deviation from strict frequency
  locking.
\end{abstract}

\maketitle

Synchronization is an intriguing phenomenon exhibited by a wide range
of physical, chemical, and biological systems~\cite{Pikovsky2001}.
The basic setting consists of coupled self-oscillating systems
synchronizing their motion; examples include such different
phenomena as orbital resonances in planetary motion or the rhythm of
muscle cells in mammal hearts. A paradigmatic and widely studied model
of synchronization is the Kuramoto model of coupled
limit-cycle oscillators~\cite{Kuramoto1984,Acebron2005}.

The most fundamental scenario of classical synchronization is the
frequency locking of a self-sustained oscillator which is externally
driven by a harmonic force~\cite{Adler1946,Pikovsky2001}.
A self-sustained oscillator takes energy from a source, e.g.~by negative
damping, and can therefore maintain stable oscillatory motion and an
undetermined phase in the presence of dissipation.  If the oscillator
is additionally driven by a harmonic force, there is a finite range of
detuning for which the oscillator is frequency-locked to the drive,
and noise can reduce or destroy this range of
synchronization~\cite{Pikovsky2001}. There are also regimes of
frequency entrainment in which the oscillator frequency is pulled
towards the drive frequency, but does not reach it.  In this case, the
frequency of the driven oscillator, the observed frequency
$\omega_{\textrm{obs}}$, differs from both the natural frequency of
the oscillator and the drive frequency.  The simplest model exhibiting
these effects is the van der Pol oscillator~\cite{Pikovsky2001} which
allows the analysis of the complex phenomenology of synchronization.

Recently, the question if synchronization exists in {\it quantum}
systems has attracted a lot of interest. There have been important
first attempts to address this problem theoretically, from communities
as diverse as trapped atomic ensembles, Josephson junctions,
and nanomechanical
systems~\cite{Shepelyansky2006,Haenggi2006,Heinrich2011,Ludwig2012,Lee2012,
Mari2013,Hriscu2013,Lee2013,Zambrini2012,Zambrini2013,Holland2013}.
Optomechanical systems~\cite{AKM2013} appear to offer a particularly
promising approach. Recent experiments have reported classical
synchronization of nanomechanical oscillators
\cite{McEuen2012,Roukes2013}, the quantum many-body dynamics of an
array of identical optomechanical cells has been predicted to show
synchronized behavior~\cite{Ludwig2012},
and quantitative measures for quantum synchronization based
on the Heisenberg uncertainty principle have been applied to
two and many coupled optomechanical cells~\cite{Mari2013}.

In this Letter, we analyze the most basic example of quantum
synchronization: a quantum version of the harmonically driven van der
Pol oscillator.  To complement other recent work which studied phase
locking~\cite{Lee2013}, we focus on frequency entrainment and frequency locking for
various detunings, driving strengths, and non-linear damping rates. We
establish the power spectrum as a theoretical and experimental tool to
characterize the observed frequency.  As one of our main results, we
find a step-like crossover to a synchronized state as a function of
the strength of the harmonic drive.  In contrast to the classical
case, there is a finite threshold value in driving.  We find that
quantum noise reduces the synchronized region as compared to the
(noiseless) classical case and leads to a deviation from strict
frequency locking.  For weak (strong) driving and for small non-linear
damping rates, frequency entrainment is reduced (enhanced) by
increasing the non-linear damping rate. Finally, we present a
realization of our model in an optomechanical setup demonstrating that
our study is directly relevant to current experimental work.

\textit{Model}.-- 
We numerically analyze a quantum version of the van der Pol oscillator
subject to an external harmonic drive. The master equation for this model
reads (in the frame rotating with the external drive)
\begin{align}\label{eqn:m1}
\frac{d\rho}{dt} = -i \left[ -\Delta \hat{b}^{\dag} \hat{b} 
+ i \Omega (\hat{b} - \hat{b}^{\dag}), \rho \right] + \gamma_{1} 
\mathcal{D}[\hat{b}^{\dag}] \rho + \gamma_{2} \mathcal{D}[\hat{b}^{2}] \rho \, ,
\end{align}
where $\Delta$ is the detuning of the drive frequency with respect to
the natural frequency of the undriven oscillator, $\Omega$ determines
the strength of the external drive, and we have set $\hbar = 1$.  The
coefficients $\gamma_{1}$ and $\gamma_{2}$ describe negative and
non-linear damping, respectively, and $\mathcal{D}$ are Lindblad
dissipators, $\mathcal{D}[O]\rho = O \rho O^{\dag} - \frac{1}{2}
\left\{ O^{\dag} O, \rho \right\}$.  We numerically calculate the
steady-state solution $\rho_{ss}$ of the master equation
Eq.~(\ref{eqn:m1}) from which we obtain the steady-state Wigner
function $W_{ss}(x,p) = \frac{1}{\pi} \int \! dy \exp(-2 i p y) \langle x+y|
\rho_{ss} | x-y \rangle$.

The corresponding equation of motion for $\beta=\langle \hat{b} \rangle$, 
\begin{align}\label{eqn:m2}
\frac{d}{dt} \beta = i \Delta \beta + \frac{\gamma_{1}}{2}\beta 
- \gamma_{2} |\beta|^{2} \beta - \Omega \, ,
\end{align}
is the equation of motion for the classical van der Pol oscillator.
Separating the dynamics of the complex variable $\beta = r e^{i\phi}$
into the dynamics of amplitude $r$ and phase $\phi$ leads to the
classical amplitude equation $ dr/dt = (\gamma_{1}/2 - \gamma_{2}
r^{2})r - \Omega \cos \phi $ and $ d\phi/dt = \Delta + (\Omega/r) \sin
\phi $ which is called Adler equation~\cite{Adler1946} if the
amplitude dynamics are negligible $(\dot{r} = 0)$.

\begin{figure}[ht]
	\centering
	\includegraphics[width=1\columnwidth]{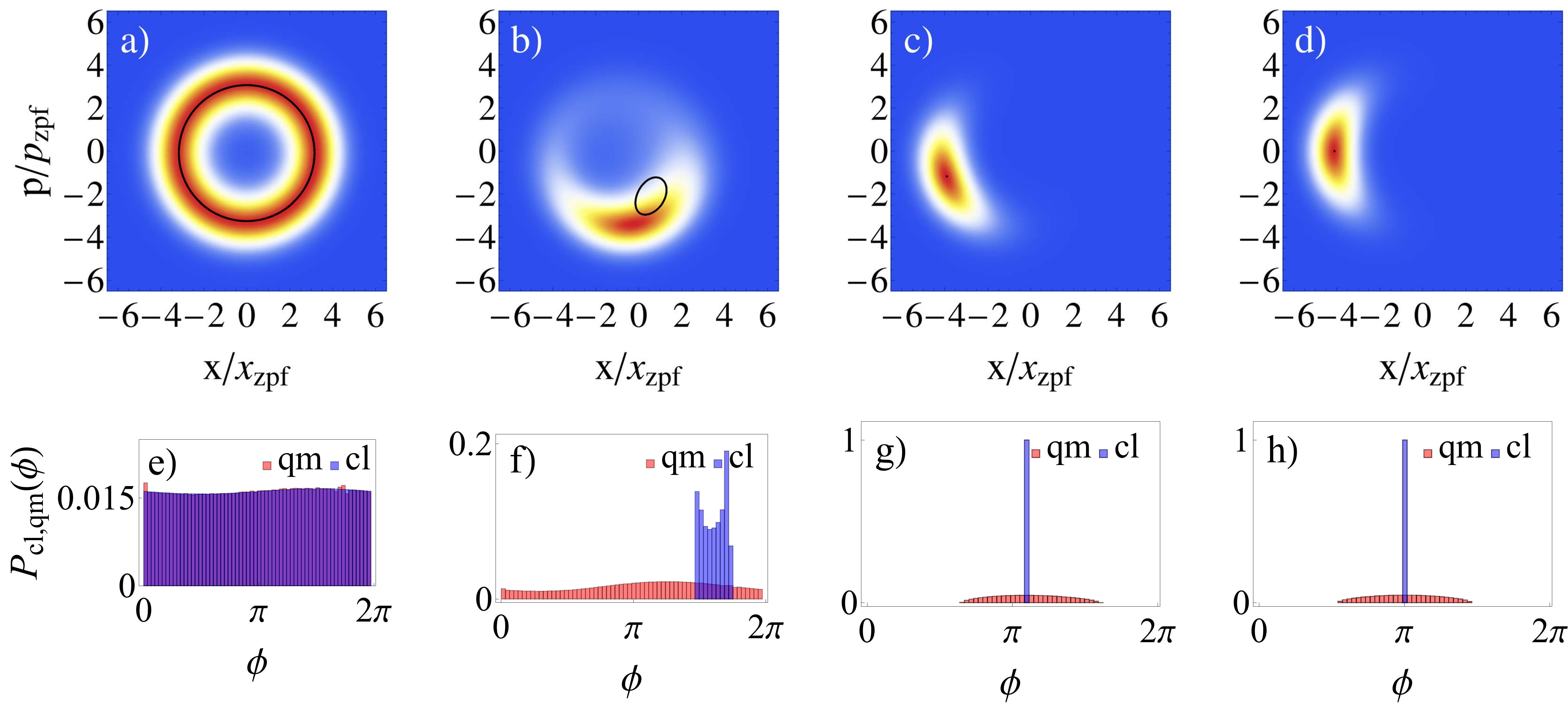}
	\caption{\label{fig:phs1} (color online). Classical phase-space trajectory
          (black solid line), Wigner function, and the corresponding
          phase probability histograms
          for $\Omega/\gamma_{1}=1$ and
          $\gamma_{2}/\gamma_{1} = 0.1$.  a), e)
          $\Delta/\gamma_{1}=16$, b), f) $\Delta/\gamma_{1}=0.6$, c),
          g) $\Delta/\gamma_{1} = 0.1$, and d), h) $\Delta/\gamma_{1}
          = 0$.
      }
\end{figure}

\begin{figure}[ht]
	\centering
	\includegraphics[width=1\columnwidth]{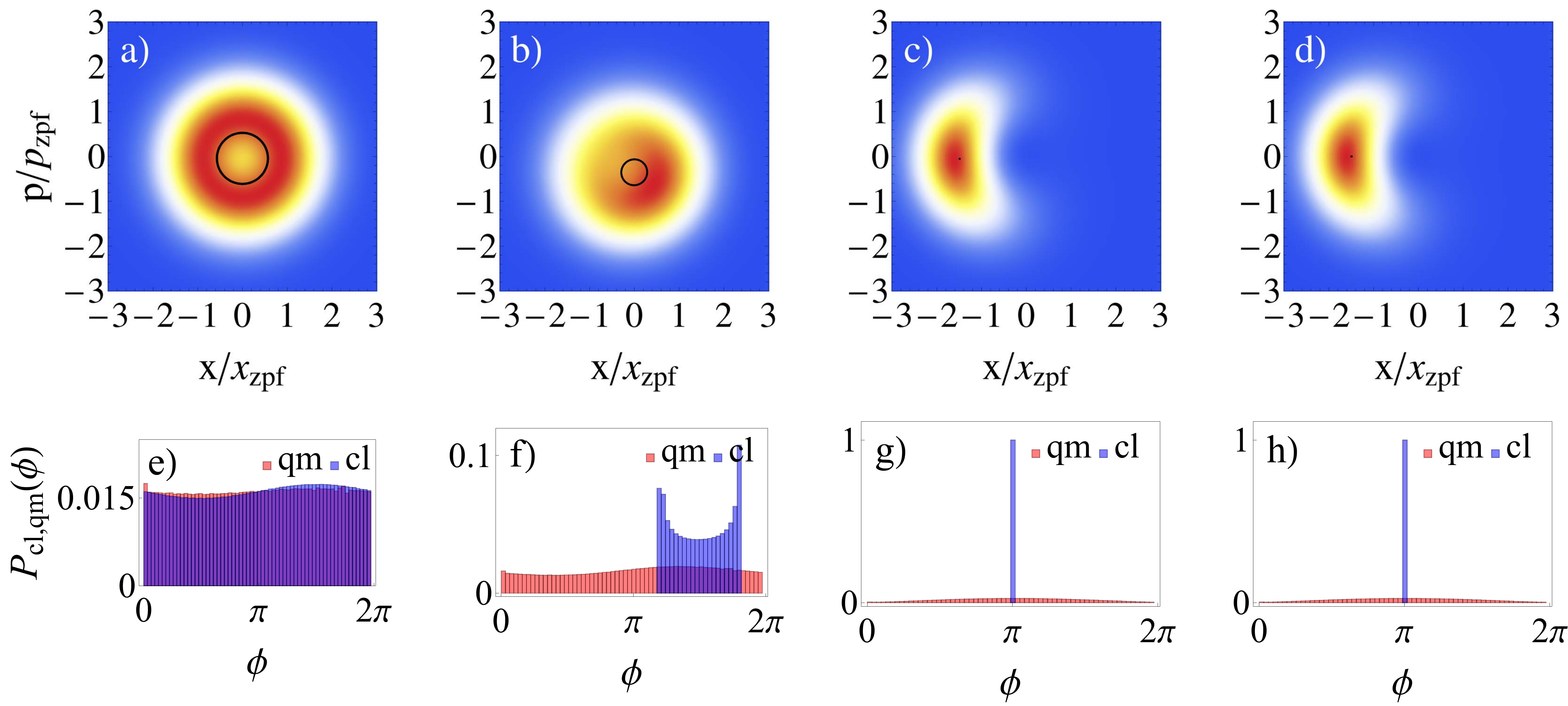}
	\caption{\label{fig:phs2} (color online). Classical phase-space trajectory
          (black solid line), Wigner function, and the corresponding
          phase probability histograms
          for $\Omega/\gamma_{1}=3$ and
          $\gamma_{2}/\gamma_{1} = 3$.  a), e)
          $\Delta/\gamma_{1}=100$, b), f) $\Delta/\gamma_{1}=12$, c),
          g) $\Delta/\gamma_{1} = 0.1$, and d), h) $\Delta/\gamma_{1}
          = 0$.}
\end{figure}

\textit{Synchronization regimes}.--
The synchronization regimes of the classical driven van der Pol
oscillator can be determined from the phase-space trajectories of its
late-time dynamics, for details on synchronization regimes and a
phase diagram of (classical) synchronization see~\cite{EPAPS}.
To characterize phase locking in the system we investigate the
probability histogram of the phase, $P(\phi)$.  In the classical case,
$P_{cl}(\phi)$ can be obtained from the late-time phase-space
trajectory. In the quantum case, we use the Wigner function, a
quasi-probability distribution for the canonically conjugate operators
$\hat{x} = x_{\textrm{zpf}} (\hat{b} + \hat{b}^{\dag})$ and $\hat{p} =
-i p_{\textrm{zpf}}(\hat{b} - \hat{b}^{\dag})$, as an analog of the
classical phase space density, where $x_{\textrm{zpf}} = 1/\sqrt{2 m
  \omega_{0}}$ is the amplitude of the zero-point fluctuations of the
oscillator with natural frequency $\omega_{0}$, and similarly
$p_{\textrm{zpf}} = \sqrt{m\omega_{0}/2}$.  Here, $P_{qm}(\phi)$ can
be calculated from $W_{ss}(x,p)$ by transforming from the variables
$x$ and $p$ to radius and phase and integrating over the radius. In
Figs.~\ref{fig:phs1} and~\ref{fig:phs2} we show the classical
phase-space trajectory, the Wigner function, and the corresponding
probability histograms for the phase variable. In both figures, we
decrease the detuning at fixed driving strength.

In the classical case, we observe phase locking, as indicated by the
$\delta$-function like shape of $P_{cl}(\phi)$ in g), h), whereas in
the quantum case $P_{qm}(\phi)$ is smeared out even in this regime
due to quantum noise. In the quantum case, a
trend towards synchronization of the quantum oscillator is visible
since the expectation values $\langle \hat{x} \rangle$ and $\langle
\hat{p} \rangle$ become finite with smaller $\Delta$, see
Figs.~\ref{fig:phs1}d) and~\ref{fig:phs2}d) where the Wigner function
concentrates in a displaced blob.

The difference between the cases $\gamma_{2}<\gamma_{1}$
(Fig.~\ref{fig:phs1}) and $\gamma_{2}>\gamma_{1}$
(Fig.~\ref{fig:phs2}) is most prominent for large detuning $\Delta$.
For $\gamma_{2}<\gamma_{1}$ the Wigner function is of ring shape and
the classical trajectory almost coincides with the maximum of the
Wigner function, Fig.~\ref{fig:phs1}a). For $\gamma_{2}>\gamma_{1}$
however, the maximum of the Wigner function and the classical
trajectory no longer coincide. In this case, the quantum nature (i.e.,
the discrete level structure) of the self-sustained oscillator becomes
more important.  In the limit $\gamma_{2} > \gamma_{1}$ the rate at
which the oscillator loses two phonons dominates over the rate at
which it gains one phonon, see Eq.~(\ref{eqn:m1}).  Thus, in steady
state for a small and moderate driving strength only the lowest Fock
states of the mode $\hat{b}$ are occupied.

\textit{Spectra and observed frequency}.-- 
To investigate frequency entrainment and locking we use the classical
and quantum-mechanical power spectrum. In the classical case, it reads
\begin{align}
S_{cl}(\omega) = \int_{-\infty}^{\infty} dt \, e^{i \omega t} \, 
\overline{\beta^*(t) \beta(0)} \, ,
\end{align}
where the bar denotes time averaging. In the quantum case, we consider
the spectrum
\begin{align}
S_{qm}(\omega) = 
\int_{-\infty}^{\infty} dt \, e^{i \omega t} \, \langle \hat{b}^{\dag}(t) \hat{b}(0)\rangle \, ,
\end{align}
where $\langle \cdot \rangle$ denotes the average with respect to the
full quantum-mechanical density matrix.

\begin{figure}
	\centering
	\includegraphics[width=0.9\columnwidth]{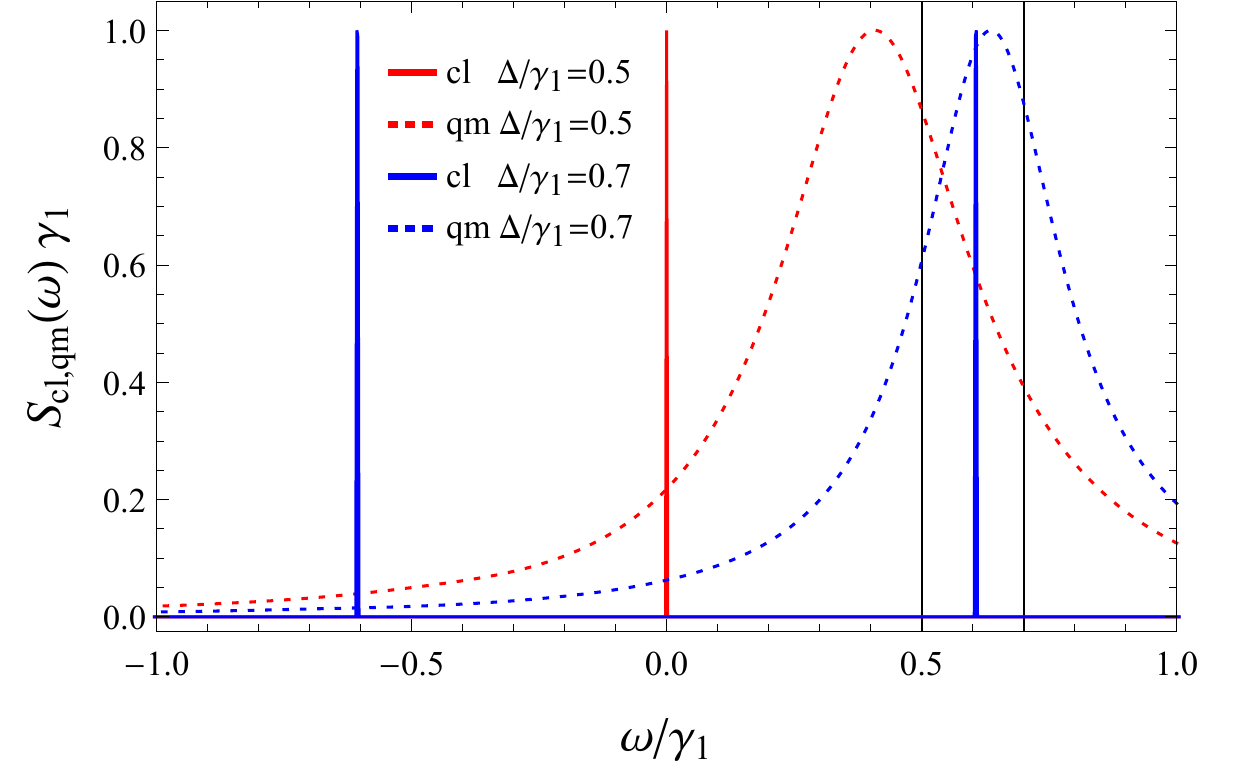}
	\caption{\label{fig:spec} (color online).
	  Normalized classical (full lines) and quantum (dashed lines) spectrum.
	  For $\Delta/\gamma_{1}=0.7$ (blue) both the quantum and classical
	  model show a peak at a frequency smaller than the detuning
	  (indicated by vertical black lines). For $\Delta/\gamma_{1}=0.5$ (red)
	  the classical model is frequency-locked to the drive, while the quantum
	  model is not. The other parameters are $\Omega/\gamma_{1}=1$ and
	  $\gamma_{2}/\gamma_{1}=0.1$.
          }
\end{figure}

Figure~\ref{fig:spec} shows typical spectra. The classical spectrum has
a $\delta$-shaped peak at the observed frequency $\omega_{\textrm{obs}}$
(in the frame rotating with the drive). In addition to this main peak, satellite
peaks emerge at higher harmonics of the main frequency \cite{Pikovsky2001}
(not shown in Fig.~\ref{fig:spec}). In the quantum case, the spectrum is
not an even function of frequency, and the peak has finite width.
In the classical and the quantum case, the observed frequency
can be obtained from the position of the maximum of $S_{cl,qm}(\omega)$
if the peaks are well-separated. The spectra are thus a convenient
tool to study synchronization properties, and we will use them in
the following to analyze the dependence of $\omega_{\textrm{obs}}$
on the detuning $\Delta$ and the driving strength $\Omega$.

\begin{figure}[ht]
	\centering
	\includegraphics[width=0.9\columnwidth]{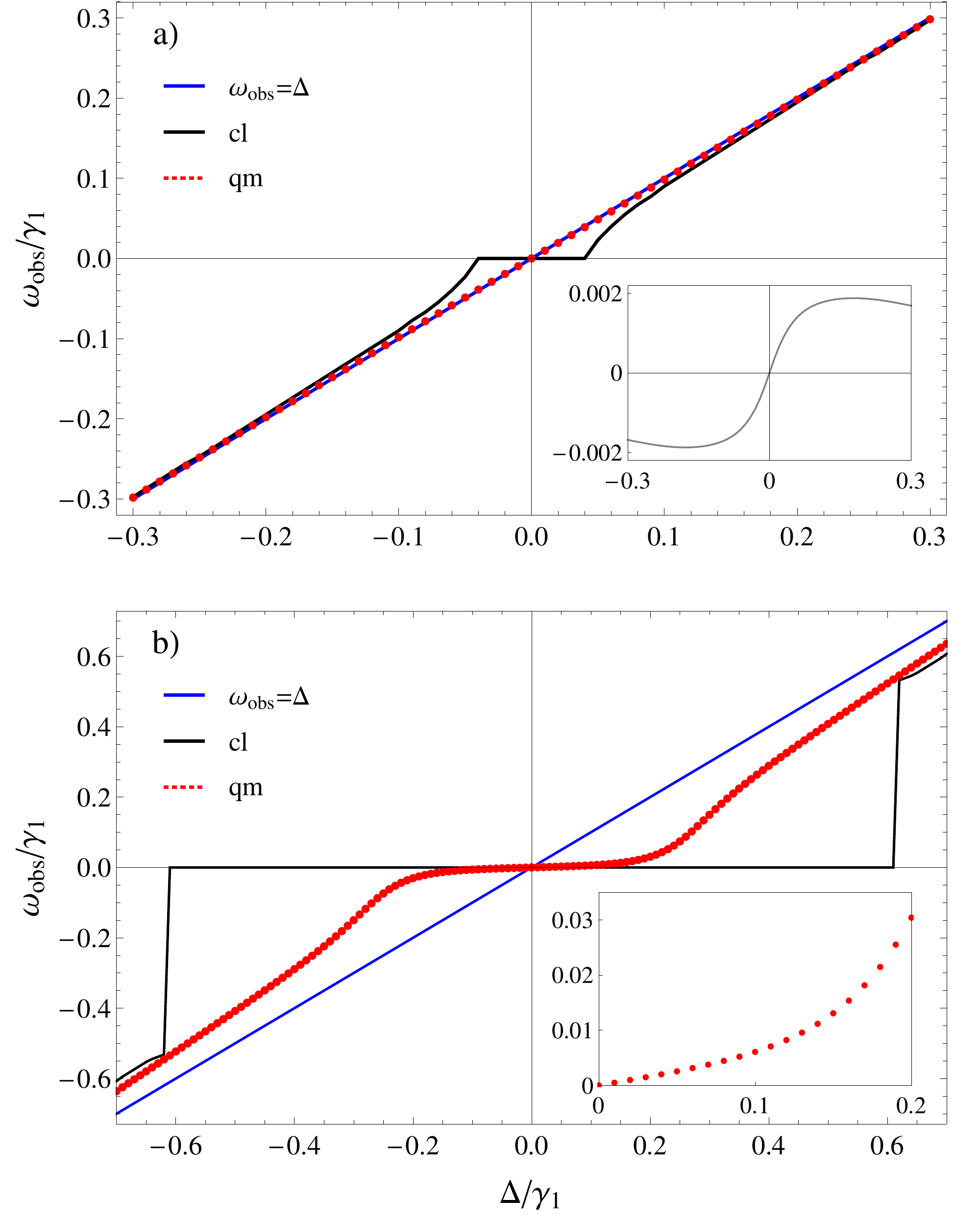}
	\caption{\label{fig:adp1} (color online).
	  Observed frequency $\omega_{\textrm{obs}}$ vs.~detuning $\Delta$ for
	  $\gamma_{2}/\gamma_{1} = 0.1$. Blue: undriven case. Black: classical
	  model. Red: quantum model.
	  In a), for $\Omega/\gamma_{1}=0.1$, the quantum model shows weak
	  frequency entrainment.
          Inset: difference between undriven and the quantum mechanical case.
          In b), for $\Omega/\gamma_{1}=1$, the quantum model shows strong
          frequency entrainment and a tendency towards frequency locking. 
          Inset: zoom-in for the quantum case at small $\Delta$.
          }
\end{figure}

\textit{Synchronization}.-- 
To investigate synchronization we plot in Figs.~\ref{fig:adp1}a)
and~\ref{fig:adp1}b) the observed frequency $\omega_{\textrm{obs}}$
as a function of detuning $\Delta$ for fixed driving strengths $\Omega$
and compare the classical case with the quantum case for
$\gamma_{2}<\gamma_{1}$.

Figure~\ref{fig:adp1}a) shows that for weak driving the classical
oscillator is frequency-locked to the drive over a finite range of $\Delta$.
In the quantum-mechanical model, however, there is only weak frequency
entrainment (cf.~inset), except at $\Delta=0$ where the frequency of
the oscillator trivially agrees with the drive.  This is in stark
contrast to the classical case, where an arbitrarily weak drive leads
to a small, but finite synchronized region. At a larger driving
strength, strong frequency entrainment and a tendency towards
frequency locking for the quantum model can be seen in
Fig.~\ref{fig:adp1}b).  However, as the inset shows, there is no exact
frequency locking at finite detuning $\Delta \not= 0$.  We attribute
the absence of frequency locking to quantum noise, since for the
classical van der Pol oscillator, noise can reduce or even destroy the
synchronization region~\cite{Pikovsky2001}. 
A semiclassical approach valid in the limit $\gamma_{1} \gg \gamma_{2}$ 
yields the appropriate (external) white noise which mimics the
(intrinsic) quantum noise and can be used to compare the quantum case
to a noisy classical van der Pol oscillator~\cite{Lee2013}.

\begin{figure}[ht]
	\centering
	\includegraphics[width=0.99\columnwidth]{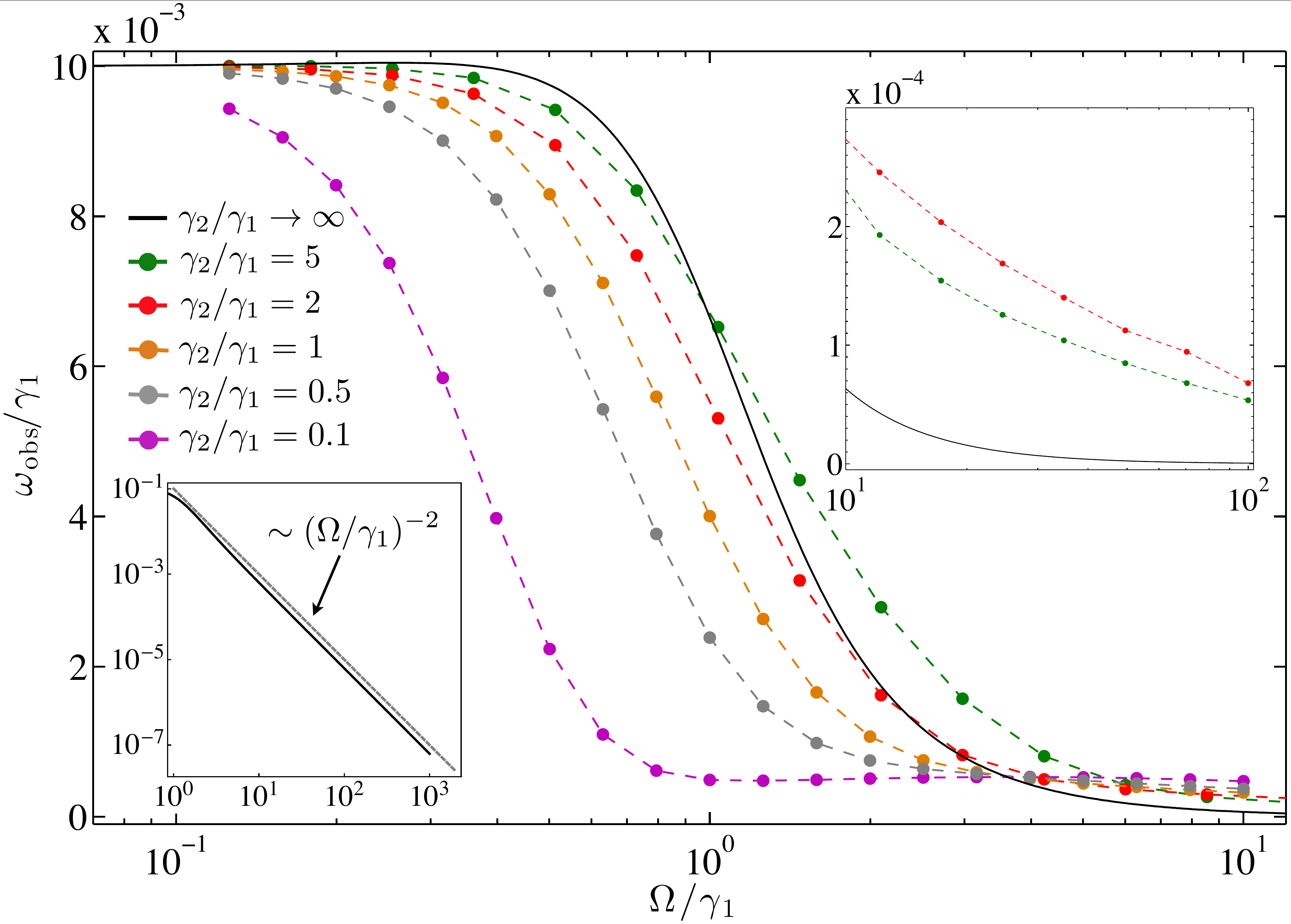}
	\caption{\label{fig:6} (color online).
	  Observed frequency $\omega_{\textrm{obs}}$ vs.~driving $\Omega$
	  in the quantum model for various values of $\gamma_{2}/\gamma_{1}$ and
	  $\Delta/\gamma_{1} = 0.01$. A step-like crossover from weak
          to strong entrainment can be observed. At larger $\gamma_{2}/\gamma_{1}$
          the step shifts to larger driving strengths. Below threshold, increasing
          $\gamma_{2}/\gamma_{1}$ reduces frequency entrainment, i.e.,
          inhibits synchronization. Right inset: Observed frequency at larger
          driving. Increasing $\gamma_{2}/\gamma_{1}$ leads to increased
          frequency entrainment, i.e., improved synchronization.
          Left inset: The observed frequency $\omega_{\textrm{obs}}$ for
          $\gamma_{2}/\gamma_{1} \rightarrow \infty$ (black)
          vanishes like $(\Omega/\gamma_{1})^{-2}$ (gray dashed).
          }
\end{figure}
\begin{figure}[ht]
	\centering
	\includegraphics[width=0.99\columnwidth]{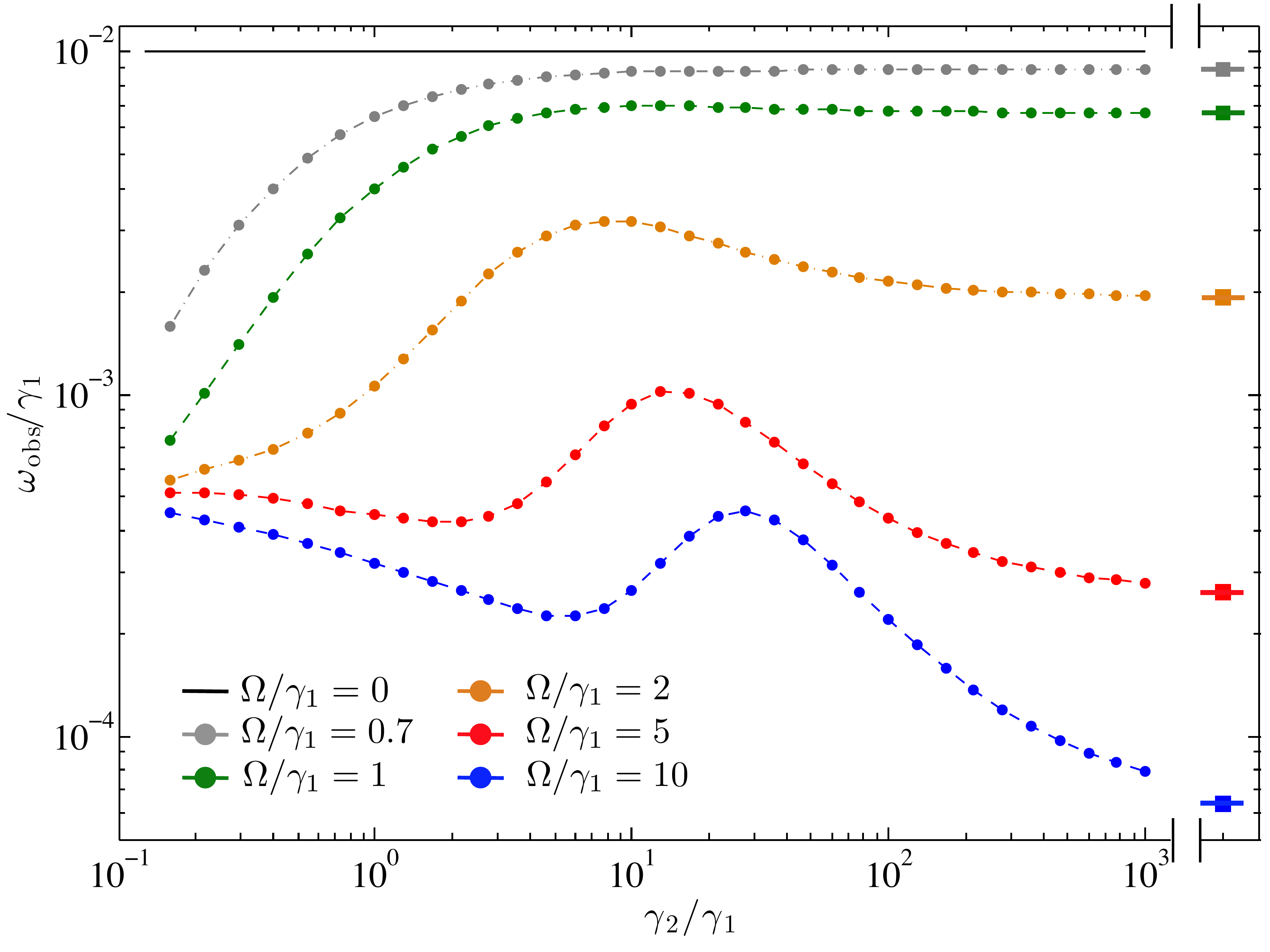}
	\caption{\label{fig:7} (color online).  Observed frequency
          $\omega_{\textrm{obs}}$ vs.~non-linear damping rate
          $\gamma_{2}$ in the quantum model for various driving
          strengths $\Omega$ and $\Delta/\gamma_{1} = 0.01$.
          Increasing the driving strength is seen to enhance frequency
          entrainment.  For weak (strong) driving and at
          $\gamma_{2}/\gamma_{1} \lesssim 1$, frequency entrainment is
          reduced (enhanced) by increasing $\gamma_{2}/\gamma_{1}$.
          Squares on the right are obtained analytically in the limit
          $\gamma_{2}/\gamma_{1} \rightarrow \infty$.  }
\end{figure}

To analyze the synchronization behavior in more detail, we now fix the
detuning $\Delta$ and study the dependence of $\omega_{\textrm{obs}}$
on the driving strength $\Omega$ and the non-linear damping rate $\gamma_{2}$.

Figure~\ref{fig:6} shows $\omega_{\textrm{obs}}(\Omega/\gamma_{1})$ for different
damping rates $\gamma_{2}/\gamma_{1}$.  A weak drive leads to a small amount of
frequency entrainment.  However, if the drive exceeds a threshold
whose position depends on the non-linear damping $\gamma_{2}/\gamma_{1}$, a
step-like crossover to strong frequency entrainment occurs.  Moreover,
the influence of $\gamma_{2}$ is different below and above the
threshold. For weak driving, increasing $\gamma_{2}/\gamma_{1}$ (and
thereby increasing the quantum nature of the oscillator) reduces
frequency entrainment, i.e., for small driving a large value of
$\gamma_{2}/\gamma_{1}$ inhibits synchronization.  For large driving,
the situation is opposite: increasing the non-linear damping rate
$\gamma_{2}$ supports frequency entrainment (see right inset of
Fig.~\ref{fig:6}).  To investigate the case of large $\gamma_{2}/\gamma_{1}$ in
more detail, we solve Eq.~(\ref{eqn:m1}) perturbatively. In the limit
$\gamma_{2}/\gamma_{1} \rightarrow \infty$ and for $\Omega/\gamma_{1}
= \Delta/\gamma_{1} = 0$ the steady state of Eq.~(\ref{eqn:m1}) can be
found: $\rho_{ss} \rightarrow \frac{2}{3} |0\rangle \langle0| + \frac{1}{3} |1\rangle
\langle1|$.  Including a finite drive and detuning leads to non-zero
off-diagonal matrix elements $\langle 1|\rho_{ss} |0 \rangle $ and
$\langle 0|\rho_{ss} |1 \rangle$.  In this case, making use of the
quantum regression theorem, we obtain an analytic expression (too
lengthy to be displayed here) for the spectrum $S_{qm}(\omega)$ which
is independent of $\gamma_{2}$. The analytic result for the observed
frequency is also shown in Fig.~\ref{fig:6}.  Although we do not find genuine
frequency-locking in the quantum case, the difference between observed
frequency and drive frequency can be made arbitrarily small.
Our analytical treatment predicts the observed frequency to vanish like
$(\Omega/\gamma_{1})^{-2}$ (see left inset of Fig.~\ref{fig:6}).

To analyze the dependence on the non-linear damping rate,
Fig.~\ref{fig:7} shows $\omega_{\textrm{obs}}(\gamma_{2}/\gamma_{1})$
for different driving strengths $\Omega$. Without driving there is no
entrainment and the observed frequency coincides with the detuning
(black line).  Driving the oscillator stronger generally enhances
frequency entrainment. However, in the left part of Fig.~\ref{fig:7},
i.e., for $\gamma_{2}/\gamma_{1} \lesssim 1$, there is an interesting
difference in the dependence on $\gamma_{2}/\gamma_{1}$. For
weak driving, frequency entrainment is reduced by increasing
$\gamma_{2}/\gamma_{1}$. For strong driving, on the other hand,
increasing $\gamma_{2}/\gamma_{1}$ enhances frequency
entrainment and synchronization (lower three curves).
In the limiting case $\gamma_{2}/\gamma_{1} \rightarrow \infty$,
$\omega_{\textrm{obs}}$ becomes independent of $\gamma_{2}$.
The observed frequency obtained from our analytical
solution is shown at the very right of Fig.~\ref{fig:7}.

\textit{Realization}.-- 
Optomechanical systems are suitable candidates to study
synchronization. We therefore describe a possible scheme to realize
the dynamics given by Eq.~(\ref{eqn:m1}) in an optomechanical
system. The two dissipative processes
$\gamma_{1}\mathcal{D}[\hat{b}^{\dag}]$ and
$\gamma_{2}\mathcal{D}[\hat{b}^{2}]$ can readily be engineered in a
``membrane-in-the-middle'' optomechanical
setup~\cite{Thompson2008}. Such a setup gives rise to a quadratic
optomechanical coupling which can be used to favor two-phonon
absorption and emission~\cite{Nunnenkamp2010}.
The dissipative terms can be created using two lasers.
The first laser is detuned to the blue one-phonon sideband and gives
rise to the negative damping term $\sim \gamma_{1}$.  The second laser
is detuned to the red two-phonon sideband leading to the non-linear
damping $\sim \gamma_{2}$. Synchronization can be probed by changing
the frequency of the harmonic drive and observing the spectrum of the
mechanical oscillator.
A different strategy based on trapped ions to realize the dynamics of
the van der Pol oscillator in a quantum system was put forward in
Ref.~\cite{Lee2013}.

\textit{Conclusion}.-- 
We have analyzed the quantum van der Pol oscillator in the presence of
an external harmonic drive. Our study establishes the power spectrum,
that is readily accessible in current nanomechanical experiments, as
an important observable of quantum synchronization.  The steady-state
Wigner function can indicate a synchronized state, but it is not
sufficient to obtain the observed frequency and thus does not allow
for studying frequency entrainment and its locking.  The information
contained in the power spectrum thus neatly complements previous
analyses of phase locking.  We have shown the existence of weak and
strong frequency entrainment in the quantum case and found a threshold
in the strength of the drive below which there is only weak frequency
entrainment between the oscillator and the drive.
Our study contributes new insights in understanding synchronization in
the quantum regime and also allows various extensions as for instance
the investigation of synchronization of two (or an ensemble of) nonidentical
quantum van der Pol oscillators. Another interesting question is whether
engineered noise \cite{Diehl2008,Verstraete2009} (classical and/or
quantum) can favor synchronization.

\textit{Acknowledgements}.-- We would like to acknowledge stimulating
discussions with S. Aldana and F. Marquardt.
This work was financially supported by the Swiss SNF and the NCCR
Quantum Science and Technology.

\textit{Note added}.- Just before submitting this manuscript we became
aware of the work of Lee and Sadeghpour~\cite{Lee2013} studying a
quantum van der Pol oscillator with a resonant weak drive.

\onecolumngrid
\setcounter{equation}{0}
\clearpage

\begin{center}
	\large{\bf{Supplemental Material for ``Quantum synchronization of a driven self-sustained oscillator''}}
\end{center}
\begin{center}
	Stefan~Walter, Andreas~Nunnenkamp, and Christoph~Bruder \\
	\emph{\small{Department of Physics, University of Basel, Klingelbergstrasse 82, CH-4056 Basel, Switzerland}}
\end{center}

\section{Details on different synchronization regimes}

\begin{figure}[ht]
	\centering
	\includegraphics[width=0.35\columnwidth]{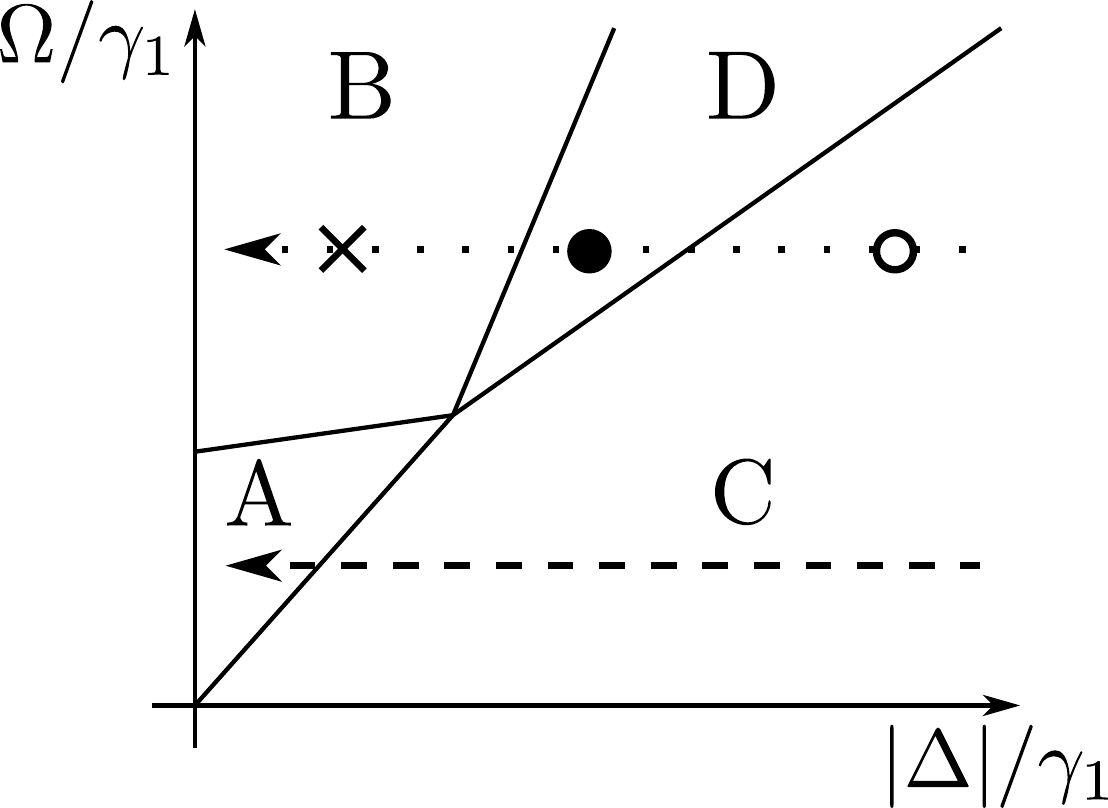}
	\caption{\label{fig:syncphases}
	Schematic phase diagram for a	harmonically driven van der Pol
	oscillator, Eq.~(2) of the main text. Regions $A$ and $B$
	correspond to a synchronized state of the classical dynamical
	system [in $A$ Eq.~(2) of the main text has three steady states
	one of which is stable, in $B$ it has one steady state which is stable].
	Region $C$ ($D$) corresponds to stable limit cycles (not) encircling
	the origin. The open and black dots correspond to the phase-space
	trajectories shown in Fig.~1a) and~1b) of the main text, and the
	cross to a point where the oscillator is synchronized with the
	drive, see e.g.~Fig.~1c) of the main text.
}
\end{figure}

Here, we give a brief overview of the different synchronization regimes.
Figure~\ref{fig:syncphases} shows a schematic phase diagram of
synchronization. At large detuning $\Delta$ the classical van der Pol
oscillator exhibits stable limit cycles with a nearly constant radius
$r_{A} = \sqrt{\gamma_{1}/(2\gamma_{2})}$ (region $C$ in Fig.~\ref{fig:syncphases}).
Following the direction of the dashed arrow in Fig.~\ref{fig:syncphases}
to smaller values of $\Delta$, the frequency of the self-sustained oscillator
is pulled towards the frequency of the drive, and at $\Delta_{A} = \Omega / r_{A}$
the oscillator synchronizes with the drive. This leads to a fixed phase relation
between self-oscillation and drive (region $A$ in Fig.~\ref{fig:syncphases}).
The region at the tail of the dotted arrow (large $\Omega$ and $\Delta$,
region C) is characterized by stable limit cycles encircling the origin.
At smaller detunings $\Delta$ (in direction of the dotted arrow in
Fig.~\ref{fig:syncphases}) the focus of the limit cycle shifts, its
radius decreases, and it no longer encircles the origin (region $D$).


\begin{thebibliography}{99}

\bibitem{Pikovsky2001}
A.\ S.\ Pikovsky, M.\ Rosenblum, and J.\ Kurths,
{\it Synchronization: A Universal Concept in Nonlinear Science}
(Cambridge University Press, New York, 2001).	

\bibitem{Kuramoto1984}
Y.\ Kuramoto, Progr. Theoret. Phys. Suppl. {\bf 79}, 223 (1984).
	
\bibitem{Acebron2005}
J.\ A.\ Acebr\'on, L.\ L.\ Bonilla, C.\ J.\ P\'erez Vicente,
F.\ Ritort, and R.\ Spigler, 
Rev. Mod. Phys. {\bf 77}, 137 (2005).

\bibitem{Adler1946}
R.\ Adler,
Proc. IRE {\bf 34}, 351 (1946);
reprinted as Proc. IEEE {\bf 61}, 1380 (1973).	

\bibitem{EPAPS}
See Supplemental Material at [link] for details.

\bibitem{Shepelyansky2006} 
O.\ V.\ Zhirov and D.\ L.\ Shepelyansky,
Eur. Phys. J. D {\bf 38}, 375 (2006).

\bibitem{Haenggi2006}
I.\ Goychuk, J.\ Casado-Pascual, M.\ Morillo, J.\ Lehmann, and P.\ H\"anggi, 
Phys. Rev. Lett. {\bf 97}, 210601 (2006).

\bibitem{Heinrich2011} 
G.\ Heinrich, M.\ Ludwig, J.\ Qian, B.\ Kubala, and F.\ Marquardt,
Phys. Rev. Lett. {\bf 107}, 043603 (2011). 

\bibitem{Ludwig2012}
M.\ Ludwig and F.\ Marquardt,
Phys. Rev. Lett. {\bf 111}, 073603 (2013).

\bibitem{Mari2013}	
A.\ Mari, A.\ Farace, N.\ Didier, V.\ Giovannetti, and R.\ Fazio,
Phys. Rev. Lett. {\bf 111}, 103605 (2013).

\bibitem{Lee2012}
T.\ E.\ Lee and M.\ C.\ Cross,
Phys. Rev. A {\bf 88}, 013834 (2013).
 
\bibitem{Hriscu2013}
A.\ M.\ Hriscu and Yu.\ V.\ Nazarov,
Phys. Rev. Lett. {\bf 110}, 097002 (2013).	

\bibitem{Lee2013}
T.\ E.\ Lee and H.\ R.\ Sadeghpour,
Phys. Rev. Lett. {\bf 111}, 234101 (2013).	

\bibitem{Zambrini2012}
G.\ L.\ Giorgi, F.\ Galve, G.\ Manzano, P.\ Colet, and R.\ Zambrini,
Phys. Rev. A {\bf 85}, 052101 (2012).

\bibitem{Zambrini2013}
G.\ Manzano, F.\ Galve, G.\ L.\ Giorgi, E.\ Hernandez-Garcia, and R.\ Zambrini,
Sci. Rep. {\bf 3}, 1439 (2013). 
	
\bibitem{Holland2013}
M.\ Xu, D.\ A.\ Tieri, E.\ C.\ Fine, J.\ K.\ Thompson, and M.\ J.\ Holland
arXiv:1307.5891 (2013).

\bibitem{AKM2013}
M.\ Aspelmeyer, T.\ J.\ Kippenberg, and F.\ Marquardt,
arXiv:1303.0733 (2013).

\bibitem{McEuen2012}
M.\ Zhang, G.\ S.\ Wiederhecker, S.\ Manipatruni, A.\ Barnard,
P.\ McEuen, and M.\ Lipson,
Phys. Rev. Lett {\bf 109}, 233906 (2012).

\bibitem{Roukes2013} 
M.\ H.\ Matheny, M.\ Grau, L.\ G.\ Villanueva, R.\ B.\ Karabalin, M.\ C.\ Cross,
and M.\ L.\ Roukes,
Phys. Rev. Lett. {\bf 112}, 014101 (2014).

\bibitem{Thompson2008}
J.\ D.\ Thompson, B.\ M.\ Zwickl, A.\ M.\ Jayich, F.\ Marquardt, 
S.\ M.\ Girvin, and J.\ G.\ E.\ Harris,
Nature {\bf 452}, 72 (2008).

\bibitem{Nunnenkamp2010} 
A.\ Nunnenkamp, K.\ B\o{}rkje, J.\ G.\ E.\ Harris, and S.\ M.\ Girvin,
Phys. Rev. A {\bf 82}, 021806(R) (2010).

\bibitem{Diehl2008}
S.\ Diehl, A.\ Micheli, A.\ Kantian, B.\ Kraus, H.\ P.\ B{\"u}chler,
and  P.\ Zoller,
Nat.\ Phys.\ {\bf 4}, 878 (2008).
	
\bibitem{Verstraete2009}
F.\ Verstraete, M.\ M.\ Wolf, and J.\ I.\ Cirac, 
Nat.\ Phys.\ {\bf 5}, 633 (2009).

\end{thebibliography}
\end{document}